\documentclass[apjl]{emulateapj}
\usepackage{hyperref}
\usepackage{color}
\hypersetup{colorlinks,
 citecolor=blue,
 linkcolor=blue}
\usepackage{soul}
\usepackage{natbib}

\usepackage{hyperref}
\hypersetup{colorlinks,%
 citecolor=blue,%
 linkcolor=blue}
\definecolor{lightblue}{rgb}{.70,.95,1} 
\def\aj{AJ}
\def\apj{ApJ}
\def\apjl{ApJ}
\def\apjs{ApJS}
\def\aap{A\&A}
\def\mnras{MNRAS}
\def\nat{Nature}
\def\pasp{PASP}%
\def\pasa{PASA}
\def\araa{ARA\&A}

\slugcomment{Draft Version \today}
\shorttitle{Tidal stirring and formation of dEs}
\shortauthors{Paudel et al.}
\usepackage[normalem]{ulem}

\begin{document}

\title{Tidal Interaction as the origin of early-type dwarf galaxies in group environment}

\author{Sanjaya Paudel\altaffilmark{1}, Chang H. Ree \altaffilmark{1}
  }
\affil{
$^1$ Korea Astronomy and Space Science Institute, Daejeon 305-348, Republic of Korea\\
}

\altaffiltext{2}{Email: sjy@kasi.re.kr}

\begin{abstract}
We present a sample of dwarf galaxies that suffer ongoing disruption by the tidal force of nearby massive galaxies. Analysing structural and stellar population properties using the archival imaging and spectroscopic data from the Sloan Digital Sky Survey (SDSS), we find that they are likely a `smoking gun' example of the formation of early-type dwarf galaxies (dEs) in the galaxy group environment through the tidal stirring. Inner cores of these galaxies are fairly intact and the observed light profiles are well fitted with the S\'ersic functions, while the tidally stretched stellar halos are prominent in the outer parts. They are all located within the 50 kpc sky-projected distance from the center of host galaxies and no dwarf galaxies have relative line-of-sight velocity larger than 205 km/s to their hosts. We derive the Composite Stellar Population (CSP) properties these galaxies by fitting the SDSS optical spectra to a multiple-burst composite stellar population model. We find that these galaxies accumulate a significant fraction of stellar mass within the last 1 Gyr, while they contain a majority stellar population of intermediate age of 2 to 4 Gyr. With these evidences, we argue that tidal stirring, particularly through the galaxy-galaxy interaction, might have an important role in the formation and evolution of dEs in the group environment, where the influence of other gas stripping mechanism might be limited. 
 
\end{abstract}


\section{Introduction}

Extreme morphology-density relations found in low-mass galaxies are considered to be the result of an effective role played by environment in the evolution of these galaxies. In particular, a class of low-mass non-star-forming galaxies, i.e., early-type dwarf galaxies (dEs), are mostly found in cluster environment and so the origin of these objects is usually connected to the environmental effects where they reside \citep{Binggeli88,Boselli06,Lisker07}.

Although variety of different mechanisms, such as ram pressure stripping \citep[RPS,][]{Gunn72}, tidal harassment \citep{Moore96} and galaxy starvation \citep{Larson80}, have been proposed to describe the way how the environment can act, but their relative effectiveness to produce the observed stellar population and structural properties of dEs remain to be understood \citep{Boselli06,Mayer01,Smith10}. RPS is expected to be more efficient in the cluster environment and evidences of ongoing ram pressure stripping in cluster environment, explicitly in Virgo cluster, are frequent in recent literatures \citep{Kenney04,Kenney14,Vollmer01}.

However, the observed diversity in stellar population and structural properties of dEs indicates that RPS alone can not explain their origin \citep{Janz14,Paudel10}. Although some features (e.g.,  presence of faint spiral arms/bars or significant rotations) may be explained with RPS scenario relating these features to a progenitor disk spiral galaxy, a significant fraction of dEs show no rotation, nor all dEs possess structural features that can be linked to the disk spiral galaxies \citep{Lisker07,Rys13}.

On the other hand, the hierarchical framework of structural growth predicts that virtually every small galaxy in the group or cluster environments is accreted from the field \citep{White78}. Such process, however, does not happen smoothly and low-mass infalling galaxies can feel a strong tidal force produced by differential gravitational acceleration which first affects the dynamics of satellite galaxies orbiting the central host potential \citep{Mayer01,Sawala12}. Known as the tidal stirring/threshing or galaxy harassment\footnote{Generally galaxy harassment is known for the cluster environment, in the group the tidal interaction between dwarf satellite and massive host galaxies in better known as tidal stirring \citep{Mayer01}. Some might call it galaxy threshing \citep{Forbes03,Koch12,Sasaki07}.}, this process not only alters the internal kinematics and structural properties of affected galaxies but also removes a significant fraction of both stellar and gas mass  \citep{Mastropietro05,Moore96}.

The role of tidal forces to shape the morphological and kinematical properties of small satellite galaxies around giant galaxies like the Milky Way has been extensively studied in recent literatures. In fact, compact galaxies,  such as M32, the  compact early-type galaxy (cE), around M31, are considered to be the survivor of these events \citep{Bekki08,Chilingarian09}. Furthermore, the state-of-art numerical simulation suggests that in both group and cluster environments, despite the act of lethal tidal force, a low-mass satellite can survive in the form of present-day dwarf-spheroidal (dSph) galaxies or dEs  \citep{Kazantzidis11,Mastropietro05,Sawala12}.

In this work, we present some of the best examples of tidal stirring that we found in the Sloan Digital Sky Survey (SDSS) imaging and spectroscopic database \citep{Ahn12,York00}, where dwarf galaxies are being disrupted by the tidal force of massive companion galaxies. They are particularly found in group environment and may provide a clue in searching for the origin of formation and evolution of dEs in group environment through the tidal interaction.

\section{Data analysis}
\subsection{Sample identification}
 As our main objective is to search for the tidal features around the dwarf galaxies in the local universe (z $<$ 0.02), we carry out an extensive search for these objects in the SDSS color images. With the inspection of the SDSS color image cut-outs, we make a catalog of all observed features, e.g., stellar tail, stream, bridge, shell and filament, around the dwarf galaxies. These features might originate from the tidal interaction with nearby massive galaxies or be the remanent of past merger activity. A complete analysis of these features with categorising their possible origin will be presented in the next series of publications. Of particular interest is the dEs\footnote{It is worth noting that we select the galaxies as dEs with visual inspection of their color image and SDSS spectroscopy. Which appear smooth -in the sense that no star-forming clumps are presence, and have no visible Balmer emission lines in optical spectrum.} which are located near to giant galaxies and prominently show up tidal features. They possess stretched outer stellar halo while the inner cores remain undisturbed and these cores are visually analog to a typical dE. Given the availability of imaging and spectroscopic data in the SDSS-III \citep{Ahn12}, we choose six candidate galaxies to study the structural and stellar population properties in detail.
  
  \begin{table}
 \caption{Sky-position and properties of Host galaxies}
 \setlength{\tabcolsep}{0.1cm}
 \begin{tabular}{lcccccl}
 \hline
name & RA & Dec &  Host & m\_r & v\_r & D\\
 & h:m:s & d:m:s & &mag(Mpc)& km/s & kpc \\
\hline
HdE1 &  00:14:37.29 &  +18:34:22.67 &    N0052 & 13.14(73) &    ---   &  18  \\
HdE2 &  00:39:15.48 &  +00:56:33.67 &    N0196 & 13.00(66) &    179  &  34  \\
HdE3 &  09:43:31.05 &  +31:58:37.01 &    N2968 & 11.47(15) &    205  &  24  \\
HdE4 &  10:22:26.03 &  +21:32:31.29 &    N3221 & 12.69(55) &    028  &  35  \\
HdE5 &  12:15:38.99 &  +33:09:35.26 &    N4203 & 11.07(16) &    024  &  37  \\ 
HdE6 &  22:37:12.42 &  +34:37:12.64 &    N7331 & 10.63(14) &    ---   &  49  \\

\hline
\end{tabular}
 \label{sample}
\,\\
For simplicity, we rename our candidate dwarf galaxies in the 1st column, and their positions in sky are given the 2nd (RA) and 3rd (Dec) columns. Name of host galaxies and their total luminosities (with distances from the Sun in Mpc) are given in the next two columns 4 and 5, respectively. The 6th column represents the difference in radial velocity between dwarf and giant companions and  the sky-projected separations between them are given in the last column.
\end{table}

 A list of basic global properties such as position, the host galaxy name and sky-projected separation from the host and relative line-of-sight radial velocity are presented in Table \ref{sample}. To convert the angular to physical separation we assume that the dwarf galaxies are at the same distance as to their hosts, which is taken from the NED\footnote{http://ned.ipac.caltech.edu} database. Interestingly, all the tidally-disrupted galaxies are found within the 100 kpc sky-projected distance from their hosts and no dwarf galaxies have relative line-of-sight velocity larger than 205 km/s to their hosts.
 
\begin{figure}
  \includegraphics[width=9cm]{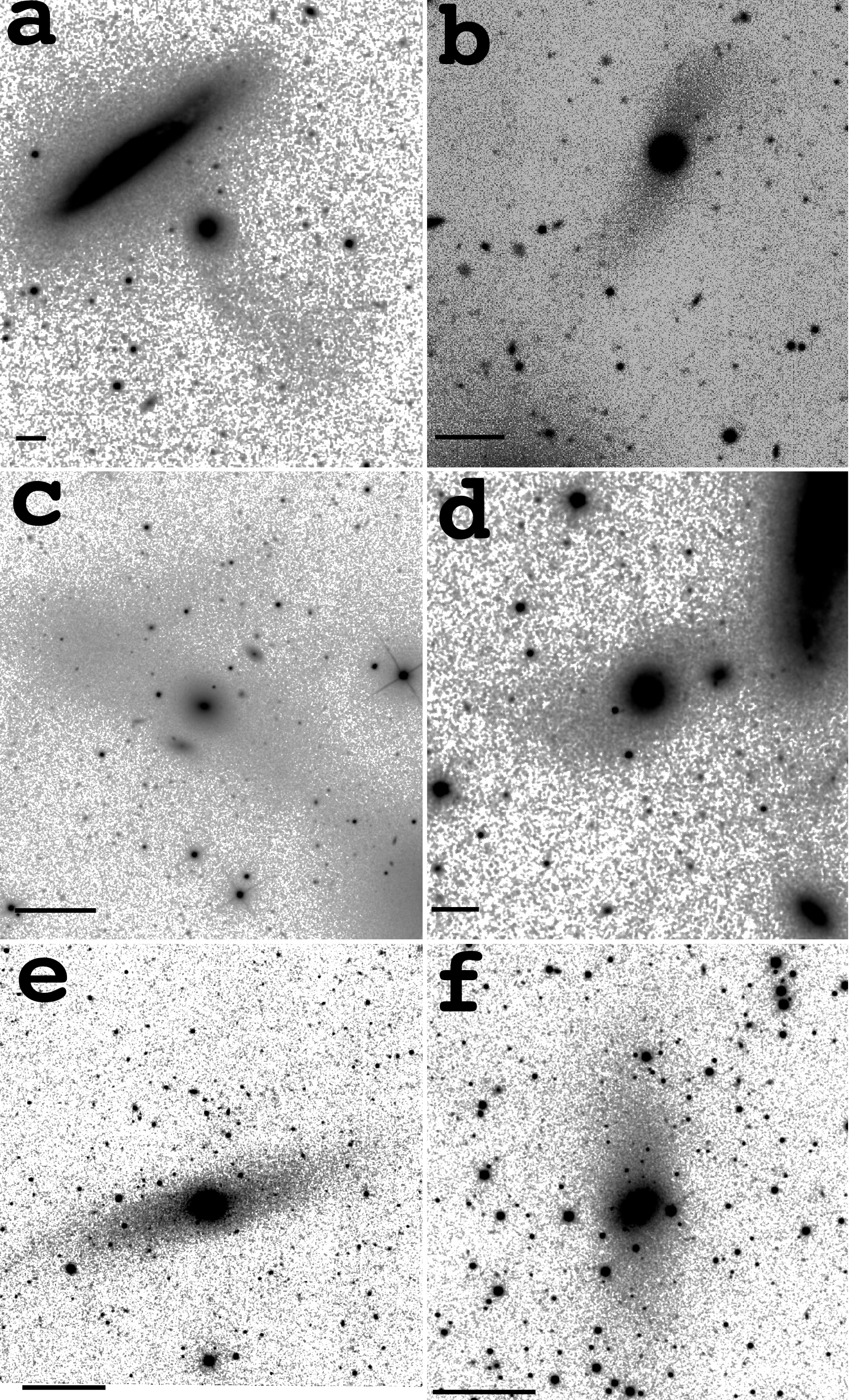}
  \caption{An optical outlook of dEs with tidal features. The images are made from the g-r-i coadded frames to gain signal at low surface brightness (LSB) component and the arcsinh profile has been applied to produce better stretching. All cut-outs have the same orientation -- the North is up and East is to the left -- and the field of view varies depending on the extension of LSB component. The horizontal black bar represents the physical scale of 5 kpc length.\\
 \hrule\vspace{0.1cm}
a:HdE1 (N0052),
b: HdE2 (N0196),
c: HdE3 (N2968),
d: HdE4 (N3221),
e: HdE5 (N4203),
f: HdE5 (N7331)
 \vspace{0.1cm}
  \hrule}
  \label{spic}
  \end{figure}

 A complete schematic view of dEs with their tidal features is shown in Figure \ref{spic}. All of them prominently show the stretching of outer stellar body while the central cores remain somewhat round. The S-shape stretching is eminent among the HdE1, HdE2 and HdE4 and the rest show the elongated stellar streams on both sides of the central body of dwarf galaxies.

 \subsection{Image analysis}
To perform the surface photometry on the optical images, we retrieve the archival images from the SDSS-III database \citep{Ahn12}. We make extensive use of $r$-band image, since it provides a higher signal-to-noise ratio (S/N) than other bands. Although the SDSS-III database provides the  sky-subtracted fits images much improved than from the previous releases, we again subtract the sky-backgrounds using the similar approach in \cite{Paudel14}. 

\begin{figure}
\includegraphics{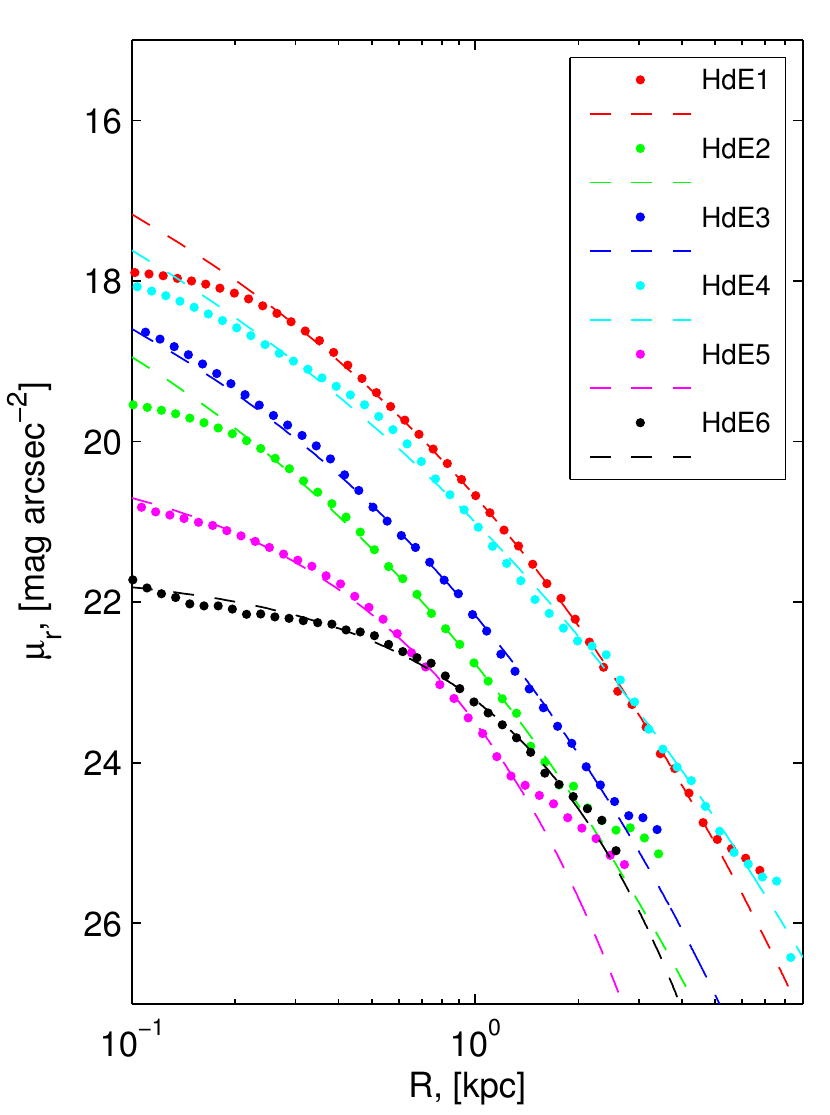}
\caption{Major axis light profile of galaxies}
\label{hprofile}
\end{figure}

IRAF $ellipse$ task has been used to extract the galaxy's major-axis light profile. Before running $ellipse$ task, we masked manually all non-related background and foreground objects including the host galaxies. Since all the candidates are well separated from the hosts we did not subtract the host galaxy light, instead we simply mask them with a sufficiently large aperture. During the ellipse fit, the center and the position angle are held fixed and the ellipticity is allowed to vary. The centers of the galaxies are calculated using the IRAF task $imcntr$ and the position angles are determined by several iterative runs of $ellipse$ before the final run. The derived one-dimensional light profile along the major axis is shown in Figure \ref{hprofile}. 

\begin{table}
\caption{Structural and photometric properties of dwarf galaxies}
\label{phot}
\begin{tabular}{lcccccr}
\hline
Name & M$_{r}$ & Re  & $<$$\mu$$>$ &  n & Re(model) & Fr\\
& (mag) & (kpc) & (mag/arc$^{2}$) & & (kpc) & \% \\
\hline
HdE1  &   -19.10   &   0.832   &   19.22 & 3.5  &    0.91 & 30\\ 
HdE2  &   -17.06   &   0.753   &   20.98 & 3.7  &    0.66 & 45\\ 
HdE3  &   -16.44   &   0.793   &   21.34 & 2.9  &    0.68 & 45\\ 
HdE4  &   -18.83   &   1.002   &   19.83 & 4.3  &    1.41 & 15\\ 
HdE5  &   -16.03   &   0.909   &   22.32 & 1.5  &    0.64 & 60\\ 
HdE6  &   -16.09   &   0.954   &   22.58 & 1.1  &    1.12 & 40\\ 
\hline
\end{tabular}
\,\\
The values in the columns 2, 3 and 4 are derived form non-parametric method, i.e., Petrosian photometry (see Text). The S\'ersic index n and Re(model) are from the best fit S\'ersic model parameters. In the last column, the light fractions of the leftover are presented after the best-fit model galaxy is subtracted from the image.
\end{table}

Using the $\chi^{2}-$minimization scheme, we fit the observed galaxy light profile to a S\'ersic function. To avoid the complexity due to the extended low surface brightness tail in the modelling of galaxy light profile, we only include the inner region during the fit. That is done with applying a cut-off on the surface brightness at 24.5 mag/sec$^{2}$. The results of the best-fit S\'ersic parameters are presented in Table \ref{phot}, and the observed and modelled one-dimensional light profiles are shown in Figure \ref{hprofile}. We find that, in some cases, relatively high S\'ersic indices for them to be considered a dE, and they are particularly the farthest one. We suspect that this is due to the inclusion of points from stretched stellar body to the fitting where a simple cut-off limit of 24.5 mag/sec$^{2}$ might not be sufficient to define the inner core of galaxy properly. 

 We measure the total light in the residual images after the model-galaxy subtraction to assess the light fraction in the stretched tidal tail. However, we warn that the derived values are not free from large uncertainty. One obvious reason for this is that the SDSS images are not deep enough for accurate surface photometry at the surface brightness level of $\sim$24 mag/sec$^{2}$.  In addition to that, the galaxy itself might not have a perfect S\'ersic profile and any residual from intrinsically imperfect galaxy model subtraction also contributes significantly.

Due to the rather complicated nature of galaxy morphology (see Fig.\ref{spic}, top panel) it is difficult to approximate the observed light distribution with a single parametric profile and then derive the global photometric parameters. The photometric parameters (e.g., half-light radius and total luminosity) are therefore derived using a non-parametric approach. We follow a similar procedure as described in \cite{Janz14,Paudel14}. The total flux is computed within an aperture of twice the Petrosian radius \citep{Petrosian76}. Since we use the ellipse output to derive the Petrosian radius and hence the total flux, the derived half-light radius is the half-light semi-major axis. We then convert it to the circularised half-light radius by multiplying a scaling factor (b/a)$^{0.5}$. Note that we do not attempt to correct for the missing flux outside of twice the Petrosian radius as suggested by \cite{Graham05}. We find that the measured half-light radius (Re)  varies in between 0.7 to 1.1 kpc. Though the statistics is small this range fairly agrees with the study of \cite{Janz14}. They measured the sizes of a large sample of early-type galaxies in Virgo cluster using the similar method, see their Figure 9.

 \subsection{Spectroscopy}
Four (HdE2, HdE3, HdE4 and HdE5) of our galaxies are targeted for spectroscopic observation in the SDSS survey. We retrieve the optical spectra of these galaxies from the SDSS data archive and they seem to have reasonable signal-to-noise ratio. They are observed with the spectrograph fed with fibers of 3{\arcsec} diameter, and since the sample galaxies cover a wide range of distances the central 3{\arcsec} represents 200 pc to 900 pc central region of galaxies depending on the distance to these galaxies. The optical spectra of these galaxies do not show up strong emission lines and resemble a typical spectrum of a non-star-forming galaxy. We find that only HdE3 contains a minute emission line of H$\alpha$ while we inspect it with zooming into this region, and not surprisingly this galaxy has the center relatively bluer than the out-skirts, similar to the blue-centered dEs in the Virgo cluster \citep{Lisker06}. 
 
  \begin{table*}
\caption{Stellar population properties}
\label{sfh}
\setlength{\tabcolsep}{4pt} 
 \begin{tabular}{l|ccc|ccc|ccr| cr}
\hline
\hline
&  Young& & Light& Intermediate&   &Light & Old&   & Light & SSP &\\
Name   &    Age(Gyr)          &     [Fe/Z](dex)            &     (\%)  &    Age(Gyr)          &      [Fe/Z](dex)            &    (\%)   &    Age(Gyr)        &      [Fe/Z](dex)            &    (\%) & Age(Gyr) & [Fe/Z](dex) \\
\hline
HdE2  &    1.32$\pm$0.04  &    0.20$\pm$0.02  &     53  &    3.59$\pm$0.39 &    -0.06$\pm$0.04 &    40  &    12.00(Fixed)  &     -1.60$\pm$0.29 &    07 &  1.8$\pm$0.2 &    0.11$\pm$0.06 \\
HdE3  &    0.32$\pm$0.02  &    0.21$\pm$0.02  &     11  &    1.46$\pm$0.02 &     0.02$\pm$0.02 &    79  &    12.00(Fixed)  &     -1.06$\pm$0.17 &    10 &  1.2$\pm$0.1 &    0.01$\pm$0.05 \\
HdE4  &    1.29$\pm$0.15  &    0.00$\pm$0.15  &     14  &    3.95$\pm$0.29 &    -0.09$\pm$0.03 &    72  &    12.00(Fixed)  &     -1.29$\pm$0.27 &    14 &  3.1$\pm$0.2 &   -0.19$\pm$0.04 \\
HdE5  &    0.18$\pm$0.02  &   -0.16$\pm$0.09  &     18  &    2.85$\pm$0.13 &    -0.25$\pm$0.04 &    51  &    12.00(Fixed)  &     -1.38$\pm$0.11 &    30 &  2.5$\pm$0.2 &   -0.44$\pm$0.05 \\

\hline
\end{tabular}
\,\\
CSPs that we decompose into the three epochs of star-formation. Young, intermediate and old populations are presented in 2nd, 3rd and 4th panel, respectively. The ages, metallicities and observed light fractions are listed in 1st, 2nd and 3rd column in each panel, respectively. SSPs parameters are presented in fifth panel.
\end{table*}

Exploiting the large wavelength coverage of the SDSS optical spectroscopy, we derive the Composite stellar population (CSP) properties of the galaxies using a full spectrum fitting method. For this purpose, a publicly available code UlySS\footnote{http://ulyss.univ-lyon1.fr} from \cite{Koleva08} has been used. We follow the procedure that is implemented in \cite{Koleva13} to study the CSP from the similar kind of spectroscopic data sets. We fit the observed galaxy spectra with a combination of three Single Stellar Population (SSP) models,  see \cite{Koleva13} for detail. The main idea is to decompose the available information in observed spectrum into three different epochs of star-formation. Although this may seem a reasonable exposition than assuming a single burst of stellar population, we should keep in mind that the degeneracy between age and metallicity becomes even more complex.

We use the SSP model from \cite{Vazdekis10}. The two (young and intermediate) ages are bounded in the range 0.1 to 0.8 and 0.8 to 5 Gyr\footnote{Note however that this is just for primary guesses. If we do not find any young component (i.e., $<$ 0.8 Gyr) we then use slightly older age boundary of 1 to 2 and 2 to 8 Gyr. Therefore the final CSPs are obtained with iterations applying different age boundaries.}. We fix the old population at an age of 12 Gyr, while the metallicities in all three episodes of star-formation are allowed to vary. Emission lines are not included in the fitted model, instead we permit the code to mask emission lines automatically using the internally imbedded clipping procedure. No corrections for the Galactic extinctions have been applied to the SDSS spectra.

The results from full spectrum fitting to derive the CSPs that we decompose into the three episodes of star-formation are listed in Table \ref{sfh}. We find the two galaxies, HdE3 and HdE5, possess a significant young components of age less than 0.5 Gyr, and other two, HdE2 and HdE4, contain no such young stellar population. It seems, however, that all galaxies in this sample accumulate a considerable fraction of stellar mass within the last Gyr. Nevertheless, the dominant populations are from intermediate ages, i.e., 2 to 5 Gyr, and they show a significant chemical enrichment during that period of star formation. We find a similar trend of ages and metallicities for all galaxies, i.e., increase of metallicity with time. Note, to some extent, that this trend is not free from so-called age-metallicity degeneracy \citep{Worthey99}.

 \section{Summary and Discussion}
We presented a caught-in-act view of the tidal disruption of dwarf galaxies around the massive galaxies. Given the large luminosity range of dE/dSph class (M$_{r}$ $\sim$ -4 to -19 mag), our candidate galaxies represent the bright end of the luminosity function of dEs in any environment (i.e., group or cluster). However, dEs of the magnitude M$_{r}$ = $\sim$ -16 are not uncommon and this luminosity range traces those which are extensively studied beyond the Local group. This is usually happen due to the extreme low surface brightness nature of these galaxies and they require extensive telescope time to get good SNR data. Our study also suffers, by design, the similar limitation as we have used the shallow SDSS imaging data. Indeed, deep imaging like NGVS  and MATLAS will allow us to explore more fainter regime \citep{Paudel13}. We therefore limit our discussion on the possible formation and evolutionary scenario for dEs, not for their faint cousins dSphs.

We find that the difference in total luminosity between the host and disrupted dwarf galaxies is always larger than 2 mag for this sample. Interestingly, the host galaxies are disk galaxies in all cases. Three of them, N0196, N2968 and
N4203, can be classified as S0s and the others are typical spiral galaxies. We find that most of them are members of well-defined groups, listed in the group catalogs of \cite{Giuricin00} and \cite{Makarov11}. N0196 is the member of a compact group \citep{Hickson89} and N3221 hosts many dwarf satellite around it. Our search in NED provides at least 8 dwarf satellites around N3221 within the 500 kpc and $\pm$300 km/s of sky-projected distance and radial velocity range, respectively, which well satisfies the group selection criteria of \cite{Makarov11}.

For those dwarf galaxies with the optical spectroscopic data in the SDSS archive, we perform a detailed study of stellar population properties. We derived CSP properties of these galaxies which can be used as a proxy for the history of stellar population build-up at three different epochs. That reveals these galaxies contain a significant fraction of young ($\sim$1 Gyr) stellar population, and in two of our sample galaxies even younger ($\sim$0.2 Gyr) stellar populations are also detected. Nevertheless, all four galaxies seem to have achieved their majority of stellar mass in the intermediate age, i.e., 2 to 5 Gyr. Since the stellar population model used for CSP study is scaled at fixed $\alpha$-abundant ratio ([$\alpha$/Fe]) equal to the solar value, the derived CSPs do not vary in [$\alpha$/Fe]. Using the method of LICK-indices, similar in \cite{Paudel10}, we derive the SSP [$\alpha$/Fe]. We find that these galaxies posses slightly sub-solar value of [$\alpha$/Fe] with an average -0.05$\pm$0.1 dex. This is consistent with the study of \cite{Paudel10} and indicates that these galaxies might have experienced relatively less intense star-formation activity in the past.

\subsection{Tidal interaction and formation of dEs}
Tidal disruption of dwarf galaxies has an important role to play in many astrophysical phenomenon. For example, outer stellar halo of giant galaxies and intra-cluster light in the cluster core are thought to be built by the stellar population of dwarf galaxies that are disrupted during the accretion into the larger system such as galaxy group and cluster \citep{Gregg98,Koch09}. Indeed recent deep imaging of nearby giant galaxies have discovered that fine structures in the form of stellar shell, filaments and streams are ubiquitous around them which primarily emerges from the disruption of dwarf galaxies by action of the tidal force from host galaxies \citep{Delgado10,Miskolczi11}. 

On the other hand, how such fundamental process shapes the properties of dwarf galaxies themselves is also a crucial issue to understand the evolution of low-mass satellite population in the group and cluster environments. At what condition these tidally-perturbed dwarf galaxies can survive and when not is critical even to solve some fundamental problems of current cosmology, such as the missing-satellite problem \citep{Henriques08,Klypin99}. Early speculation of formation of compact satellite, such as M32-type galaxies, by tidal stripping \citep{Faber73} has been nearly well-established by the recent observations of unequivocal evidences \citep{Huxor11}. But how the same tidal force influences the evolution of majority of satellite galaxies that are of dEs/dSphs class, is still under debate.

As these iconic examples are suggesting, galaxy stirring might be more important to form dEs particularly in group environment than other gas-stripping mechanisms. With the detailed study of a sample of dEs using the 2-dimensional kinematic information \cite{Rys14} claim that, even in the Virgo cluster, the tidal harassment might be more responsible to produce the present-days kinematic and structural properties of dEs.

The comparison between the observed stellar population ages and some form of dynamical time scale of interaction might provide a better understating of how the interactions were unfolded and how they evolved in the past. A simple speculation of ages of the interactions is not trivial as the nature of the problem \citep[see][]{Lokas13}. However, as \cite{Lokas13} predicts the probability of detection of tidal debris around the satellite dwarf is maximum near to the pericenter passage, so we expect that our dwarf galaxies are located not far from the pericenters of their orbits.

While this work presented here is certainly not a conclusive understanding of formation and evolution of dEs in group environment, we have shown the `smoking-gun' examples of tidal stirring phenomenon that may help not only in the understanding of formation and evolution of dEs but also to decipher the process of interaction of galaxies itself. Further detailed studies of the internal kinematics of these galaxies might scrutinize our findings.

The study is entirely based on SDSS archival data (http: //www.sdss.org/) and has made use of NASA Astrophysics Data System Bibliographic Services and the NASA/IPAC Extragalactic Database (NED).

\end{document}